# Oxygen-vacancy centers in $Y_3Al_5O_{12}$ garnet crystals: electron paramagnetic resonance and dielectric spectroscopy study

V. Laguta[1], M. Buryi[1], S. Tkachenko[2], P. Arhipov[2], I. Gerasymov[2], O. Sidletskiy[2], O. Laguta[3], M. Nikl[1]

[1]*Institute of Physics Academy of Science of Czech Republic, Cukrovarnická 10, 162 00 Prague 6, Czech Republic*
[2]*Institute for Scintillation Materials, National Academy of Sciences of Ukraine, Nauky av. 60, 61001 Kharkiv, Ukraine*
[3]*Institute of Physical Chemistry, University of Stuttgart, Pfaffenwaldring 55, D-70569 Stuttgart, Germany*

**Abstract**

$F^+$ center, an electron trapped at oxygen vacancy ($V_O$), was investigated in the oxygen deficient $Y_3Al_5O_{12}$ (YAG) garnet crystals by electron paramagnetic resonance (EPR). The measurements were performed in the wide temperature interval 5–450 K and the frequency region 9.4–350 GHz with using both the conventional continue wave and pulse EPR technique. The pulse electron-nuclear double resonance was applied to resolve the hyperfine interaction of the trapped electron with surrounding nuclei. The measurements show that at low temperatures, T < 50 K, EPR spectrum of the $F^+$ center is anisotropic with g factors in the range 1.999–1.988 and originates from three magnetically inequivalent positions of the center in garnet lattice according to different directions of the Al(IV) – $V_O$ – Al(VI) chains, where Al(IV) and Al(VI) are the tetrahedral and octahedral Al sites, respectively. As the temperature increases, the EPR spectrum becomes isotropic suggesting a motional averaging of the anisotropy due to motion of $F^+$-center electron between neighboring oxygen vacancies. With further increase of the temperature to T > 200 K, we observed delocalization of the $F^+$-center electron into the conduction band with the activation energy about 0.4–0.5 eV that resulted in substantial narrowing of the EPR spectral line with simultaneous change of its shape from the Gaussian to Lorentzian due to diminish up to zero of the Fermi contact hyperfine field at $^{27}Al$ and $^{89}Y$ nuclei. Such temperature behavior of the $F^+$-center electron in YAG is completely similar to behavior of a donor electron in a semiconductor. Our findings is further supported by measurements of the conductivity and dielectric properties. In particular, these data show that the conduction electrons are not homogeneously distributed in the crystal: there are high-conductive regions separated by poorly-conductive dielectric layers. This leads to the so-called Maxwell-Wagner dielectric relaxation with huge apparent dielectric constant at low frequencies. To the best of our knowledge, this is probably the first observation of a donor-like behavior of $F^+$ center in wide band-gap insulating crystals.

## 1. Introduction

$Y_3Al_5O_{12}$ (YAG) and isomorph $Lu_3Al_5O_{12}$ (LuAG) crystals as well as their ceramics and powders doped with rare-earth ions are widely exploited as laser medium [1,2,3,4], luminescent and scintillation materials in the high-tech industry, medicine and security imaging and monitoring systems [5,6,7], and in the solid state white light sources [8]. In spite of the favorable scintillation properties, the main demerit of both YAG and LuAG is the presence of slow components in the scintillation decay, which causes serious degradation of the light yield and timing characteristics [9,10]. It is commonly accepted that these slower decay components are related to host defects which temporarily trap charge carriers before their radiative recombination at emission centers. The luminescence efficiency of the garnets is also undergone an influence of the host defects. Among different possible defects, so-called antisite ion defects, i.e. Y or Lu at the Al site and vice versa and oxygen vacancies are the most frequently mentioned intrinsic defects in garnet crystals [11,12,13]. Moreover, both these defects can be coupled into pairs as it was demonstrated in $YAlO_3$ [14]. Oxygen vacancy ($V_O$) naturally serves as an effective trap for electrons. It can be thus filled by one or two electrons forming $F^+$ and $F^0$ centers, respectively. In spite that F type centers usually produce absorption band in the visible or UV region, identification of these centers by only optical methods is quite questionable, especially in non-cubic complex oxides where the actual local structure of the center can hardly be determined from only optical data. In this respect, the $F^+$ centers, being paramagnetic, can be successfully identified and studied at atomistic level by electron paramagnetic resonance (EPR) as it was demonstrated in many oxide materials [15,6,16,17,18].

The probably first indication on EPR detection of an electron trapped at $V_O$ in YAG was published by Mori [19]. A single EPR line at the g factor 1.995±0.002 was observed in the additively colored crystal. This EPR line correlated with appearance of three broad absorption bands peaked at $1.2\times10^4$, $2.0\times10^4$, and $2.8\times10^4$ cm$^{-1}$. It was assumed that the observed EPR line as well as the absorption bands originate from an unpaired electron trapped at an oxygen vacancy ($F^+$ center). However, no further studies were performed with this EPR signal as it was uninformative to make valid conclusion about origin, local structure and thermal stability of the corresponding center. Later, similar spectrum (an isotropic line at g factor 1.996) was revealed in YAG crystals either γ-irradiated or photo-irradiated by UV nitrogen laser at 77 K [20]. In the crystals reduced at 1500-1650 K a single EPR line at g=1.994 with the small anisotropy, Δg=0.0002 was also measured [21]. Again, excepting the fact of this EPR signal observation and possible $F^+$ center origin, no further analysis of this spectrum was done. Strong EPR signal with g factor 1.994 was observed in YAG crystal doped by Si [22]. This crystal was slightly colored in blue. The spectrum in this crystal was attributed to an electron trapped at $V_O$ nearby Si ion, i.e. $F^+$ - Si complex. This center was stable at room temperature and existed without any prior irradiation. Note that a weak spectral line with g factor 1.994 was also observed in high-quality transparent YAG and LuAG crystals, but it was difficult to interpret its origin as it appeared only after X-ray irradiation at liquid nitrogen temperature together with the spectrum from trapped hole centers and these two spectra are imposed one to another [23].

There is also an uncertainty in the temperature stability of the oxygen vacancy center in YAG and, respectively, in the interpretation of corresponding thermally stimulated luminescent peaks, absorption and luminescence bands (see, e.g. [20,24,25], review paper [6]) which appear in crystals after X-ray and UV irradiation at 77 and 300 K. Some of the mentioned peaks may also arise from the antisite defects created by Y cations occupying Al sites and vice versa. They serve as effective traps for electrons as well. Moreover, the Y antisite ion can be coupled with oxygen vacancy creating the $Y_{Al}^{3+} - V_O$ defect acting as electron trap. Such defects were identified by EPR in YAlO$_3$, for instance [14]. But EPR spectra of these defects were never observed in YAG or LuAG, in spite that luminescence and TSL data indicated on possible existence of such defect in garnets as well [10,24,26,27].

In this paper we present results of detailed EPR investigation of colored in blue YAG crystals which surely contained high concentration of oxygen vacancies. For comparison, annealed in air YAG crystal was measured as well. To overcome saturation effects in EPR spectra due to extremely long spin-lattice relaxation time of the oxygen vacancy center and to resolve anisotropy in g factor majority low-temperature measurements were performed at Q microwave band at 34 GHz using the spin-echo detected EPR [28]. Moreover, some spectra were measured at frequencies up to 350 GHz. To resolve the hyperfine interaction of the trapped electron at oxygen vacancy with its surroundings pulse electron-nuclear double resonance (ENDOR) was applied in the Q microwave band. A model of F$^+$ center in YAG is proposed on the base of the obtained experimental dat. The model assumes localization of an electron within oxygen vacancy space at T < 50 K, like in MgO, and its further behavior with temperature increase similar as for donor electron in semiconducting material.

## 2. Experiments

The crystals were grown by the Czochralski method in pure reducing atmosphere. They were colored from light to deep blue color. These crystals become completely transparent after annealing in air atmosphere. EPR measurements were performed using Bruker E580 spectrometer operated at X (9.4 GHz) and Q (34 GHz) microwave (MW) bands in the both conventional continue wave (CW) and pulse modes at the temperatures from 450 K down to 5 K. In addition, to resolve hyperfine interaction in the EPR signals, pulse Electron Nuclear Double Resonance technique was utilized as well. The ENDOR spectra were measured at Q band with using the Mims [29] and Davies pulse sequences [30].

Additional EPR measurements were performed with the home-made high frequency spectrometer operated at frequencies from 82.5 GHz up to 1100 GHz in the magnetic field 0 – 15 T and the temperature 4.6 and 260 K. The field sweep resolution was 28000 points for the 0– 15 T region with the sweep time of 100-150 min for one spectrum. Details of this spectrometer design can be found in [31] and in supplementary material to this publication.

### 3. Experimental results

#### 3.1. CW EPR data

EPR spectra measured at 175 K in the colored (a) and transparent (b) crystals are shown in Fig. 1, as an example. The colored sample contains a strong signal at ≈ 337 mT, g factor 1.994 and several weaker signals attributed to well known $Fe^{3+}$ spectrum in YAG crystals [32]. As a background impurity, these ions always present in YAG or LuAG crystals. It is shown below that the strong spectral line at g=1.994 originates from an electron trapped at oxygen vacancy, usually called $F^+$ center [15]. The $F^+$ signal completely disappears after annealing of the colored crystal in air at 1000 $^0$C for about 6 hours confirming that this signal is indeed related to oxygen vacancies. Note that $Fe^{3+}$ concentration also decreases with this annealing due to recharge of iron ions [21]. Similar spectrum at g factor about 1.994 was measured by us in YAG doped by Si [22] and in the X-ray irradiated YAG and LuAG crystals doped by Sc and Mg [6], and Mori in YAG crystal annealed in Al atmosphere [19].

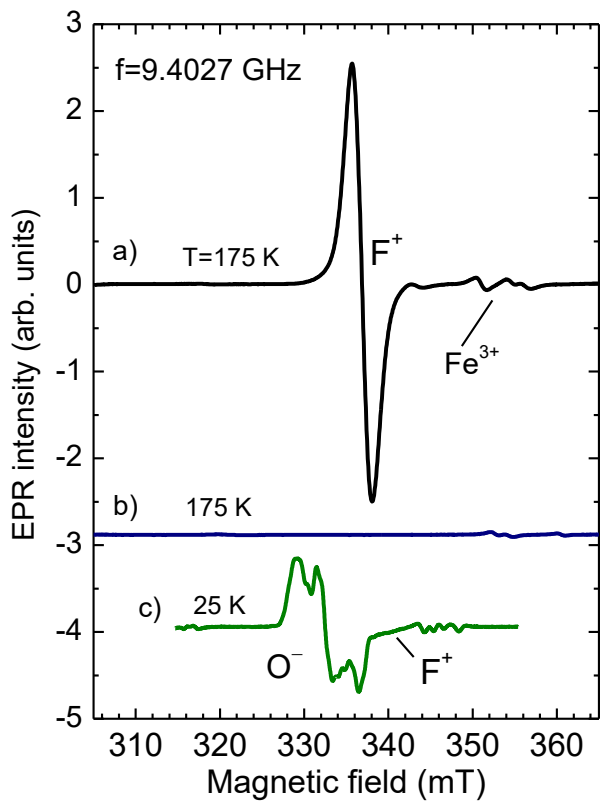

Fig. 1. X-band EPR in YAG crystals before (a) and (b) after annealing in air atmosphere. The strong signal at ≈ 337 mT is produced by $F^+$ center; other weak intensity signals are due to $Fe^{3+}$ impurity ions. (c) For comparison, X-ray induced spectrum in YAG doped by 500 ppm Mg is shown as well. The F$^+$ signal appears as a shoulder at the right side of the strong O$^-$ hole spectrum (for details, see [23]).

The F$^+$ center in YAG has several characteristic features. For instance, in contrast to similar oxygen vacancy center in YAlO$_3$ [14], the F$^+$ spectrum in YAG can be measured even at room temperatures and without any irradiation. Its spectral line is completely isotropic at these temperatures and g factor only slightly deviates from the g factor of free electron (g – g$_s$ = 0.0083). Below, this center is considered in details.

The F$^+$ spectrum could be measured by CW EPR technique to about 50 K. At lower temperatures, due to long spin-lattice relaxation time, spectrum decreases in intensity and becomes distorted in shape. To obtain the true undistorted spectrum at T < 50 K it was measured by the spin-echo detected field sweep method [28]. Other fact to which we pay attention, is that the spectral linewidth decreases with temperature increase that is unusual feature, usually indicating fast motion of an electron between inequivalent positions near a defect in a lattice. When this motion becomes slow at low temperature, shape of the spectral line becomes asymmetric or the line even splits into several components according to number of magneto inequivalent positions of a center [33,34].

In order to distinguish if the spectrum is anisotropic at low temperatures, its temperature dependence was measured in the Q band where due to four-time larger frequency the separation between components in the spectrum should increase as well. Fig. 2 presents such spectra taken at few temperatures to demonstrate temperature behavior of the F$^+$ spectrum. The temperature dependence of the full width at haft maximum (FWHM) of the spectral line measured at two MW frequencies is presented in the inset to Fig. 2. In particular, one can see that the spectrum substantially increases in width with temperature decrease and finally at T < ~ 40 K its shape becomes markedly asymmetric. This asymmetric line can be fitted by three Gaussian lines (Fig. 3). The difference in the g factors of the two outermost lines, g$_3$ – g$_1$ = 0.011, approximately characterizes anisotropy of the g factor, which is small as compared to the F$^+$ center in YAP, for example [14], where the g factor anisotropy is g$_3$ – g$_1$ = 0.17. Obviously, the three-component spectrum originates from three magnetically inequivalent positions of the center in lattice while there are 96 positions of the oxygen vacancy in the unit cell which, however, transform one into another by symmetry operations of the $Ia3d(O_h^{10})$ space group of the YAG lattice [35]. These magnetically inequivalent positions of F$^+$ center correspond to different directions of the Al(IV) – V$_O$ – Al(VI) chains (paths) in the unit cell. The anisotropy of the F$^+$ center in YAG is determined by involving of the Al and Y *p* and *d* orbitals into the predominantly s-type ground state of the F$^+$ center. This is further supported by data of ENDOR measurements.

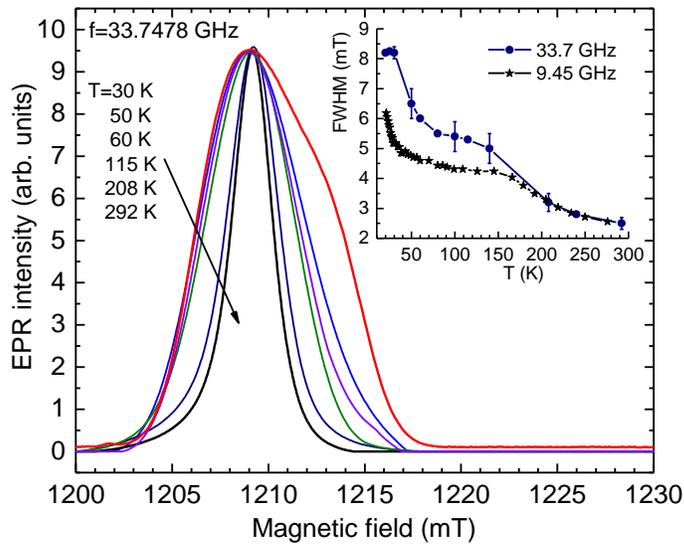

Fig. 2. Q band EPR absorption spectra in YAG crystal at different temperatures normalized in intensity to the spectrum at 292 K. The spectrum at 30 K (red line) was obtained by spin echo detected field-sweep method. Other spectra were measured by a conventional CW EPR technique with integration of the original EPR signals. The inset shows temperature dependence of the FWHM linewidth of the spectral lines measured at 9.45 and 33.7 GHz.

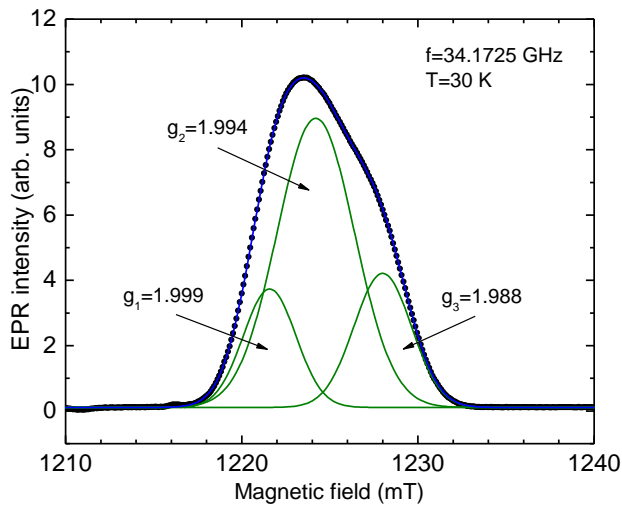

Fig. 3. Decomposition of the EDEPR line into three Gaussian components demonstrating anisotropy of g factor of the $F^+$ center at low temperatures, T < ~ 40 K.

### 3.2. High frequency EPR measurements

The $F^+$ center was further studies by EPR measurements at high-frequencies from 90 GHz up to 350 GHz. Even at these frequencies there is only one spectral line from the $F^+$ center at room temperatures, which only slightly increases in width as the frequency increases but does not split. For instance, the FWHM linewidth is only 3.5-4.0 mT at the frequency 250 GHz (Fig. 4, upper spectrum), while the expected splitting for the static spectrum (slow motion regime) due to g factor anisotropy depending on crystal orientation should be up to 45 mT at this frequency. Therefore, even at this high frequency the spectrum is still motinally averaged at room temperatures (fast motion regime). Contrary, it becomes clearly split into few components at low temperature (Fig. 4, bottom spectrum). Note that at these high MW frequencies, the $Fe^{3+}$ central transition spectrum is superimposed on the $F^+$ spectral lines that makes problematic correct separation of the $F^+$ resonances at most crystal orientations as well as their dependence on temperature and frequency.

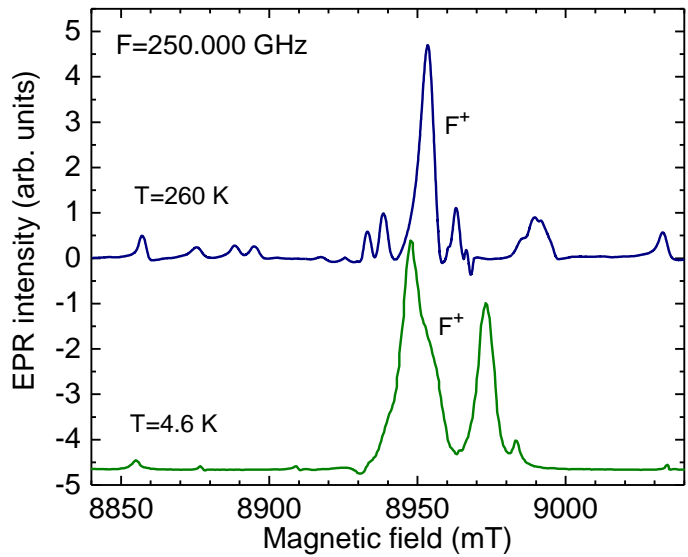

Fig. 4. EPR spectrum of $F^+$ center in colored YAG crystal measured at the frequency 250 GHz at 260 K (upper spectrum, fast motion regime) and 4.6 K (bottom spectrum, slow motion regime). Other low-intensity lines belong to $Fe^{3+}$ resonances. The crystal has arbitrary orientation with respect to magnetic field.

### 3.3. ENDOR data

Characteristic feature of any $F^+$ center is delocalization of electron density over surrounding cations that leads to Fermi contact hyperfine (HF) interaction of the electron spin with nuclear magnetic moments. The corresponding HF splitting was not resolved in the EPR spectrum due to small splitting between HF components. Therefore, the HF interaction was measured by ENDOR technique. The ENDOR spectrum measured by the Mims pulse sequence [29] (Davies pulse sequence gives the same spectrum, but of lower intensity) is shown in Fig. 5.

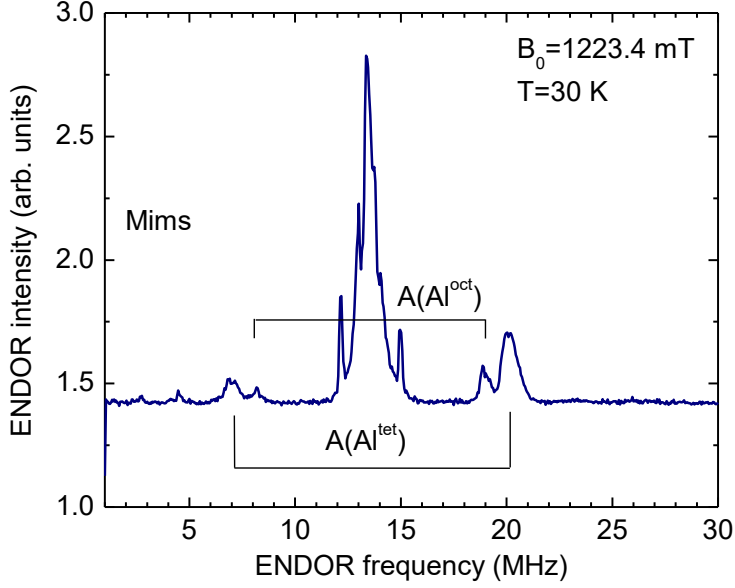

Fig. 5. ENDOR spectrum for the F$^+$ center measured by the Mims pulse sequences.

The spectrum contains group of strong lines centered at the Larmor frequency 13.63 MHz of the $^{27}$Al isotope which has the nuclear spin I = 5/2 and natural abundance 100%. The splitting between outermost lines in the group is about 3 MHz. Other well visible lines grouped into two duplets are separated by much larger frequency distances, 13.2 and 10.7 MHz (A(Al$^{oct}$) and A(Al$^{tet}$) in Fig. 5). Positions of these two duplets do not depend markedly on the crystal orientation, while the lines near the $^{27}$Al Larmor frequency are substantially changed in both intensity and frequency positions being however within the 3.6 MHz frequency interval. This suggests that this group of lines originates from distant to paramagnetic center $^{27}$Al nuclei and the complex ENDOR structure around the $^{27}$Al Larmor frequency originates mainly from quadrupole transitions of the $^{27}$Al nuclei.

The observed ENDOR spectrum can be interpreted using the following spin Hamiltonian:

$$\mathbf{H} = \beta \mathbf{B} g \mathbf{S} - g_n \beta_n \mathbf{B} \mathbf{I} + \sum_i \left\{ S A_i I_i + \frac{\nu_Q}{2} \left[ 3 I_{iz}^2 - I_i(I_i+1) + \frac{1}{2}\eta \left(I_{i+}^2 - I_{i-}^2\right) \right] \right\}, \qquad (1)$$

where $\nu_Q = \dfrac{3e^2qQ}{h2I(2I-1)}$ is the quadrupole frequency, $eQ$ is the quadrupole moment of a nucleus. The axes x,y,z are the principal axes of the tensor **V** that describes the electric field gradient (EFG) with $|V_{zz}| \geq |V_{xx}| \geq |V_{yy}|$, $eQ = V_{zz}$, $\eta = \dfrac{V_{xx} - V_{yy}}{V_{zz}}$. In the spin Hamiltonian (1), the first two terms describe the electron and nuclear Zeeman interactions, the third term is the electron-nuclear HF interaction and the last term describes the quadrupole interaction of the nuclear spins $I_i$ with the electric field gradient.

Because for distant $^{27}$Al nuclei the HF interaction is negligibly small as compared to the quadrupole one, the HF term in the spin Hamiltonian (1) can be omitted and the ENDOR resonances will thus depend exclusively on the EFG value. For the tetragonal symmetry of the EFG tensor and the condition $g_n\beta_n BI \gg \nu_Q$ that is valid for both the tetrahedral and octahedral Al sites in the field of 1.2 T, the solution of the spin Hamiltonian (1) neglecting second order effects is simple:

$$\nu_m = \nu_L - \nu_Q(m-1/2)\frac{3\cos^2\theta - 1}{2}. \qquad (2)$$

Here $\nu_L = \beta_n g_n B$ and $\theta$ is the angle between **B** and the direction of the $z$ axis of the EFG tensor. This axis in YAG coincides with the direction of the tetrahedron and octahedron distortion, [100] and [111] crystal axes, respectively. Taking $\theta = 0$ that corresponds to maximum splitting between quadrupole transitions, we found that the splitting between the 5/2 ↔ 3/2 and -5/2 ↔ -3/2 transitions will be $4\nu_Q$. The quadrupole frequency $\nu_Q$ for $^{27}$Al is known from NMR measurements [36]. It is 0.915 and 0.0955 MHz for the tetrahedral and octahedral Al sites, respectively. This agrees well with the positions of the quadrupole satellites measured by ENDOR taking into account that the actual orientation of the [100] crystal axis is not exactly along magnetic field, i.e., the angle $\theta$ in Eq. (2) is only close to zero. In particular, due to ten times larger quadrupole frequency for the tetrahedral Al sites, only quadrupole transitions from these Al nuclei are resolved in the ENDOR spectrum. Besides, the spectrum contains broad background line which originates from contribution of Al nuclei located at a middle distances from oxygen vacancy. For these nuclei, EFG is distributed in both the value and direction of the principal axes that leads to essential broadening of the quadrupole transitions. Moreover, HF interaction can not be neglected for these nuclei as well. It will also be distributed in value. Only nuclei situated enough far from the oxygen vacancy give sharp narrow spectral lines.

The HF term in the spin Hamiltonian (1) can not be neglected for the neighbor Al nuclei. Moreover, it seems that the EFG at Al nuclei in the nearest vicinity of the oxygen vacancy is so large that only the 1/2 ↔ -1/2 central transition is seen in the ENDOR spectrum. For the central transition in the first order approximation the resonance frequencies are determined by the following expression [37]:

$$\nu_{1/2} = |\nu_L \pm A_i/2|. \qquad (3)$$

Here we also assumed an isotropic HF interaction in accordance with our experiment (two duplets in Fig. 5 practically do not depend on crystal orientation). Using Exp. (3) and the data in Fig. 5, the following HF constants were determined for two Al nuclei: $A_1 = 10.7$ MHz and $A_2 = 13.2$ MHz. Obviously, the larger HF constant describes interaction of the trapped electron in F$^+$ center with Al nucleus arranged in the tetrahedral coordination due to shorter O – Al$^{tet}$ distance (0.1761 nm in regular lattice [35]) as compared to the O – A$^{oct}$ distance (0.1937 nm in regular lattice). This assignment of the ENDOR peaks is further supported by the fact that number of Al tetrahedral positions is 1.5 times bigger than for the octahedral Al. However, main reason of the markedly different intensities is related to relaxation times which should by different for these two Al sites, as even in the regular YAG lattice EFG in tetrahedron is ten times larger than in octahedron. Namely the quadrupole interaction effectively shortens relaxation time which, in turn, influences the ENDOR intensity.

Electron trapped at oxygen vacancy must also interacts with $^{89}$Y nuclei which have the nuclear spin 1/2, natural abundance 100% but very small Larmor frequency, 2.57 MHz even at the field 1.2 T. These nuclei may be responsible for two weak lines at 2.7 and 4.5 MHz. However, this assignment is not convincing as most probably the $^{89}$Y Larmor frequency and HF constant are comparable in value. As a result, not all resonances are visible due to the limitation in ENDOR frequencies to 1 MHz. The measurements in much more stronger magnetic fields can clarify the situation with the $^{89}$Y ENDOR. On the other hand, namely Al ions play major role in the interaction of trapped electron with its surroundings that is reflected in large HF constants for Al nuclei.

### 3.4. Model of the F$^+$ center in YAG and its behavior with temperature

The above presented data clearly suggest that the EPR signal with the nearly isotropic g factor in the range 1.999 – 1.988 in YAG crystals belongs to an electron trapped at oxygen vacancy (Fig. 6). Since the HF interaction of the trapped electron with surrounding nuclei is predominantly isotropic, Fermi contact type, wave function of the trapped electron is nearby spherical s-type. This corresponds to classical F$^+$ center (similar to F$^+$ center in MgO [15] or Al$_2$O$_3$ [38], for instance), when the trapped electron is mainly located within the vacancy space, but its wave function has non-zero density at surrounding cations [15]. The delocalization of electron density leads to a partial contribution of the Y 4d and Al 3p orbitals into the F$^+$ center ground state that explains the small negative shift of its g factor.

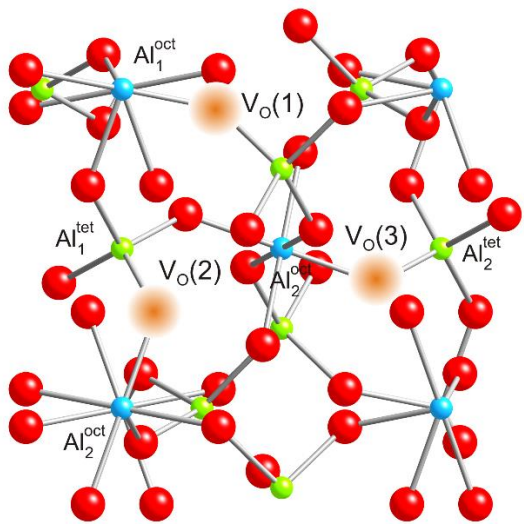

Fig. 6. Fragment of the YAG crystal structure (projection into (001) plane) with an oxygen vacancy center at three magnetically inequivalent positions.

The most interesting feature of the oxygen vacancy center in YAG is the essential narrowing of its spectral line with temperature increase, which suggests possible motion in the center. Let us discuss this phenomenon, at least, qualitatively. Fig. 7 shows the temperature dependence of the peak-to-peak linewidth (distance between two peaks of the first derivative of spectral line) measured up to 450 K at 9.4 GHz. This simply measured parameter of spectral line is commonly used for characterization of lines with different shapes, especially when the shape of a line changes with temperature. In our case, the spectral line is complex at T < 50 K (Fig. 3). It becomes Gaussian between 50–180 K, and changes from the Gaussian to Lorentzian at 150–350 K (Fig. 7b).

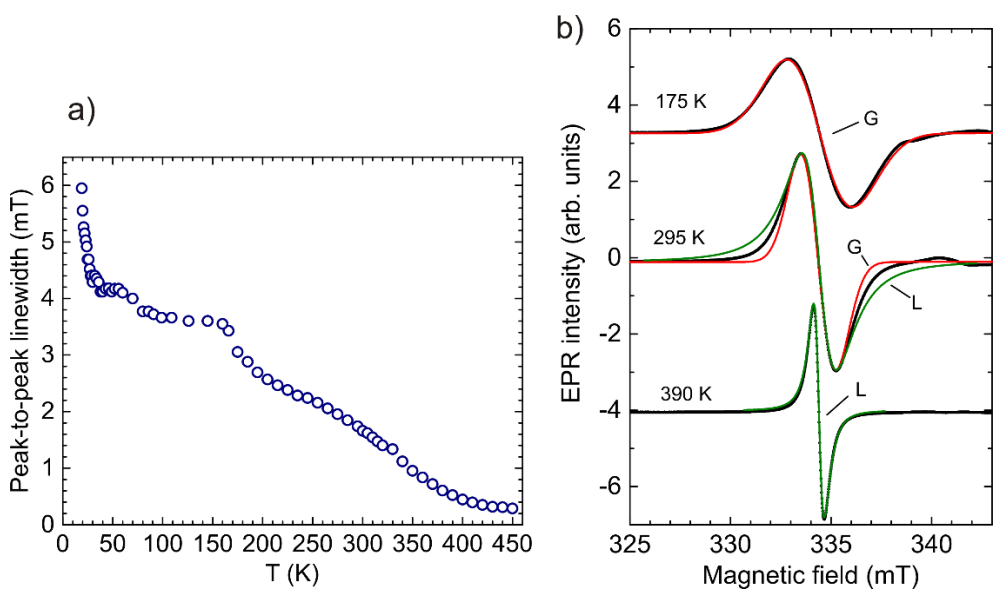

Fig. 7. (a) Temperature dependence of the peak-to-peak linewidth for F$^+$ center in YAG at 9.4 GHz; (b) simulation of line shape at different temperatures: black points are experimental data, green and red solid lines are Lorentzian and Gaussian, respectively.

One can see two temperature regions where the line substantially narrows. In the first temperature region, it sharply decreases from about 6 mT to ~3.5 mT as the temperature increases to 100 K. The g factor at this temperature is $g_0 = \frac{1}{3}(g_1 + g_2 + g_3) = 1.994$. The narrowing of the spectrum can be explained in the following way. At the lowest temperature, where the F$^+$ center shows g factor anisotropy, electron is well localized at oxygen vacancy. When the temperature increases, the electron can thermally jump to other oxygen vacancy, not filled by an electron ($V_{O1}^{\cdot} + V_{O2} \rightarrow V_{O1} + V_{O2}^{\cdot}$ process), or there could be rather thermally assisted electron tunneling between oxygen vacancies. We can not also exclude the situation of electron motion (exchange) between F$^+$ and F centers ($V_{O1}^{\cdot\cdot} + V_{O2}^{\cdot} \rightarrow V_{O1}^{\cdot} + V_{O2}^{\cdot\cdot}$ process) that will lead to averaging of the g factor anisotropy as well. Of course, such electron motion is possible when the distance between oxygen vacancies is not too large, at least is few lattice constants or when the electron wave function is quite delocalized. The motion mechanism will decrease the linewidth according to the well know relation [33]:

$$\Delta B_M = \gamma_e (\delta B_0)^2 \tau, \qquad (4)$$

where $\delta B_0$ is half of the expected splitting in the spectrum due to g factor anisotropy (when the motion is frozen), $\gamma_e$ is the electron gyromagnetic ratio and $\tau$ is the relaxation time for charge hopping. This expression is valid at the condition $\gamma_e(\delta B_0)\tau \ll 1$, i.e. at the fast motion regime. For thermally activated process, the relaxation time $\tau$ may be expressed in the form of the Arrhenius law:

$$\tau = \tau_0 \exp\left(\frac{E_a}{kT}\right), \qquad (5)$$

here $\tau_0$ is the high-temperature relaxation time and $E_a$ is the activation energy. For thermally assisted tunneling mechanism expression for $\tau$ is more complex. But it contains the exponent too (see, e.g. [39]). The fit of the experimental linewidths at T < 100 K is shown in Fig. 8. Parameters of the fit are $E_a$ = 1.2 meV and $\tau_0 \approx 5\times10^{-16}$ s. Such low activation energy suggests tunneling mechanism as the thermal depth of the F$^+$ center should be much bigger. Note that this our finding is supported by observation of a background emission in thermally stimulated luminescence (TSL) of YAG/LuAG:Ce crystals at low temperatures, which only slightly depends on temperature. To explain this fact, a tunneling (or thermally assisted tunneling) of electrons to Ce ions in the garnet lattice was proposed [10]. The concentration of F$^+$ centers as determined from EPR intensity is about $5\times10^{18}$ spins/cm$^3$ that is comparable with the common occurrence Ce concentration.

The linewidth is almost constant between 100–180 K. The spectral line at these temperatures is Gaussian. It describes well taking into account HF Fermi contact interaction of electron spin with nuclear spins of two neighboring Al ions with the HF constants determined from ENDOR measurements, $A_1 = 10.7$ MHz and $A_2 = 13.2$ MHz. Each of the Al HF components is additionally broadened by a weak HF interaction with two neighboring $^{89}$Y nuclei and more distance $^{27}$Al nuclei. The simulated spectrum for T = 175 K is shown in Fig. 7b, as an example.

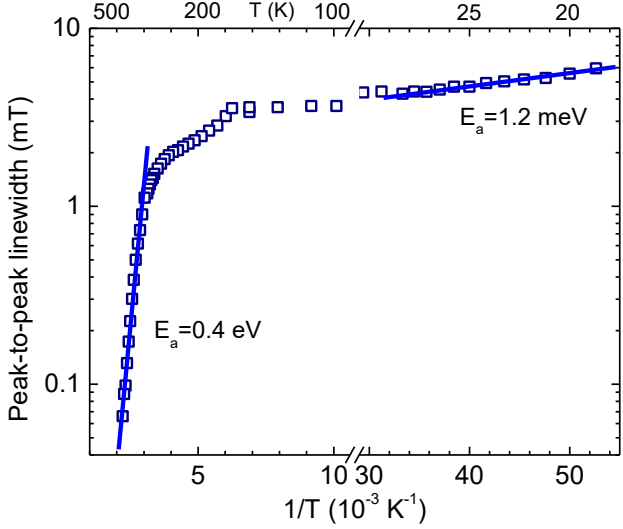

Fig. 8. Logarithmic plot of the peak-to-peak linewidth for F$^+$ center as a function of the reciprocal temperature.

It is completely surprising that the F$^+$ EPR spectrum continues to narrow at further increase of temperature in spite that both g factor and HF interaction are isotropic at these temperatures and the linewidth cannot be smaller than that determined by HF interaction. Moreover, spectral line of trapped electron usually broadens at such high temperatures due to increase of the spin-lattice relaxation rate. The narrowing of the F$^+$ spectral line indicates that the Fermi contact interaction of the F$^+$ center electron with neighboring nuclei reduces in value with temperature increase. The Fermi contact interaction is directly proportional to electron density at nucleus,

$$A = \tfrac{8}{3}\beta\beta_N g g_N |\psi(0)|^2, \tag{6}$$

where $\beta$ and $\beta_N$ are the Bohr and nuclear magnetons, $g_N$ is the nuclear g factor, and $\psi(0)$ is the electronic wave function at nucleus. Therefore, according to the relation (6) substantial delocalization of the F$^+$-center electron can exist at T > ~250 K, i.e. at these temperatures the F$^+$-electron becomes free conduction electron, like for a donor electron in n-type semiconductors [40]. This is because at T > 250 K, the linewidth follows exponential dependence on the reciprocal temperature with the activation energy ≈ 0.4 eV (Fig. 8). This energy can be considered as a very rough approximation of the electron binding energy. From this point of view, the F$^+$ center in YAP can be associated with a donor center as its temperature behavior is very similar to behavior of a donor center in a semiconductor [40].

The increased radius of wave function will also lead to exchange interaction between electron spins. The fast spin exchange with the exchange frequency $\omega_{ex}$ decreases EPR spectrum as well leading to so-called exchange narrowed line with the Lorentzian shape and width [41]:

$$\Delta B_{ex} = \gamma_e (\delta B_0)^2 / \omega_{ex}. \qquad (7)$$

Allowing temperature dependence of the exchange frequency (determined by the exchange integral and thus wave functions), one can obtain additional effective source for spectrum narrowing.

The conduction electrons can cause conductivity in the crystal. However, no direct-current (*dc*) conductivity was detected in the YAG crystals with F$^+$ centers. Instead, the crystals show marked alternating-current (*ac*) conductivity, which exponentially depends on the reciprocal temperature (Fig. 9). This suggests that conduction electrons do not travel across whole crystal due to inhomogeneity of crystal, when more conductive regions are separated by poorly conducting layers. For instance, the poorly conducting layers could be located near dislocations.

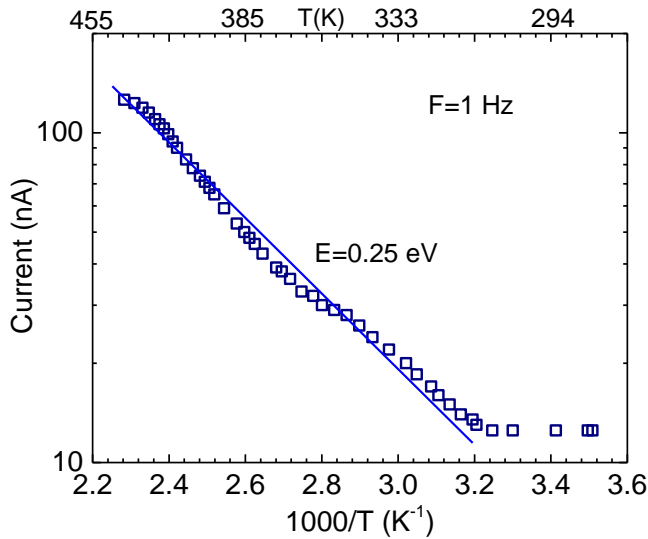

Fig. 9. Logarithmic plot of the ac current at applied ac field of 20 V/cm at the frequency 1 Hz as a function of the reciprocal temperature.

To check such a possibility, we measured frequency dependence of crystal capacitance recalculated in dielectric constant. It shows typical behavior (Fig. 10) as for a dielectric material which contains high-conductive regions isolated (or separated) by low-conductive thin dielectric layers, so-called Maxwell-Wagner dielectric relaxation [42,43]. It predicts the following dispersion for the real part of dielectric constant of inhomogeneous semiconductor or dielectric [44]:

$$\varepsilon' = \varepsilon_\infty + \frac{\varepsilon_0^p - \varepsilon_1}{1 + \tau_\varepsilon^2 \omega^2}, \quad (8a)$$

$$\varepsilon_0^p = \varepsilon_2 \frac{x\rho_1^2 + \rho_2^2}{(x\rho_1 + \rho_2)^2}, \quad (8b)$$

$$\tau_\varepsilon = \varepsilon_0 \varepsilon_2 \frac{\rho_1 \rho_2}{x\rho_1 + \rho_2}, \quad (8c)$$

where $\rho_1$, $\rho_2$, $\varepsilon_1$, and $\varepsilon_2$ denote the resistivity and the dielectric constant of the high-resistivity layers and the low-resistivity regions, respectively; $\varepsilon_0 = 8.85 \times 10^{-12}$ F·m$^{-1}$, and $\omega = 2\pi f$ is the ac field frequency. It is also assumed that the ratio of the thickness of the high-resistivity layers to the thickness of the low-resistivity regions is $x \ll 1$, and that $\varepsilon_1 = \varepsilon_2$, and $\rho_1 \gg \rho_2$.

Taking the dielectric permittivity of YAG $\varepsilon_2 = 11.7$, the calculated curve from Eqs. (8) predicts well all characteristic features of the dielectric constant with change of frequency (solid line in Fig. 10): the dielectric constant sharply increases at f < 10$^6$ Hz and approaches the dielectric permittivity of YAG $\varepsilon_2 = 11.7$ at f > 10$^7$ Hz. The following parameters were determined from the fit: $\varepsilon_0^p = 5370$, and $\tau_\varepsilon = 1.3 \times 10^{-6}$ s$^{-1}$. Unfortunately, Eqs. (8) do not allow determination of the $\rho_1$, $\rho_2$ and $x$ parameters separately as the $x\rho_1$ can be comparable with $\rho_2$ value. Therefore, more experimental data are needed for determination of all parameters of the model that is beyond of this paper.

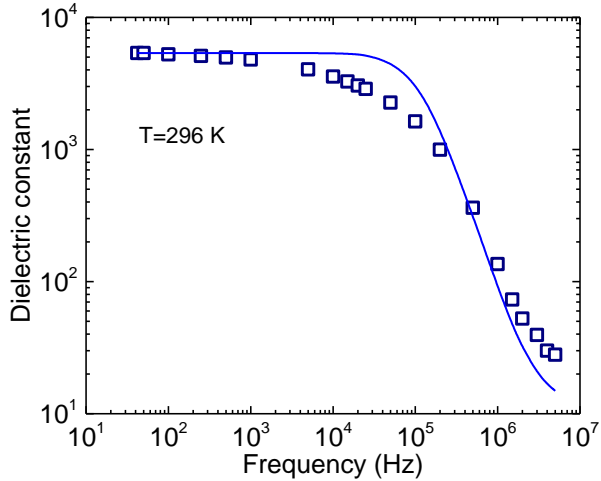

Fig. 10. Frequency dependence of the apparent dielectric constant at T = 295 K.

Let us estimate the electron activation energy from the *ac* current measurements. Considering the low-temperature limit $E_a/kT \gg 1$, the conduction electron concentration will varies with temperature as $\exp(-E_a/2kT)$. Then neglecting temperature dependence of electron mobility we can take the same functional dependence for current as for the conduction electron concentration. It gives $E_a = 0.5$ eV, the approximate value of the $F^+$- center donor level. It agrees with the value $\approx 0.4$ eV determined from EPR. The life time of the $F^+$-center electron according to Eq. (4) is $\sim 10^{-8}$ s at room temperature. The $E_a = 0.4$-$0.5$ eV energy is considerably smaller as compared with that in simple oxides MgO and CaO [15], ZnO [45,46], $Al_2O_3$ [38]). For instance, the $F^+$ center in $Al_2O_3$ is thermally stable to the annealing temperature of 700 K [38]. Another fact to which we want to pay attention is that in complex oxides with two types of cation ions as a rule the oxygen vacancy serves only as a perturbation for electron localization at one of neighboring cation ions. In such a case, the oxygen vacancy is most probably filled by two electrons being a charge neutral defect. Such oxygen-vacancy center can be usually created in oxygen deficient crystals only by UV or X-ray irradiation at cryogenic temperature (at or below 77 K) as it is shallow. Its temperature stability is below room temperature. These centers are characterized by g factors of the cation ion with are substantially shifted from the g factor of free electron. For instance, the $Pb^+$ - $V_O$ center in $PbWO_4$ has the g factors between 1.22 and 1.61 [47]. Other examples are $BaTiO_3$ [48], $PbTiO_3$ [49], $CaWO_4$ [50], $Y_2SiO_5$ [18]. The mechanism of electron localization for such oxygen-vacancy center is mainly related to polaronic effect. The center can be thus considered as a small polaron bound to vacancy. From this point of view, the $F^+$ center in YAG is probably the first example of the oxygen-vacancy center in complex oxides where an electron is located predominantly at oxygen vacancy and only weakly participates in covalent bonding with surrounding cations.

Finally, in conclusion note that the presented here results for $F^+$ center in YAG are not specific ones for only our crystals. Similar EPR spectral line from $F^+$ center was measured in YAG crystal reduced at 1500-1650 K [21], colored YAG doped by Si [22], YAG crystals either γ-irradiated or photo-irradiated by UV nitrogen laser [20] and in additively colored crystals [19]. However, no detailed studies of the $F^+$ center behavior with temperature were performed in these early publications.

**Acknowledgements**

The authors gratefully acknowledge the financial support of the Czech Science Foundation (project No. 17-09933S), the Ministry of Education, Youth and Sports of Czech Republic under project LO1409 and CZ.02.1.01/0.0/0.0/16_013/0001406. We also thank J. van Slageren (Stuttgart) for providing access to the high-frequency EPR spectrometer.